\documentclass[epj-ND]{svjour}
\usepackage[tbtags]{amsmath}
\usepackage{amssymb}
\usepackage{graphicx}

\usepackage{txfonts}

\begin{document}

\title{MCHIT - Monte Carlo model for proton and heavy-ion therapy}

\author{Igor Pshenichnov  \inst{1,2} \fnmsep \thanks{Presenting author, \email{pshenich@fias.uni-frankfurt.de}}
   \and Igor Mishustin  \inst{1,3} 
   \and Walter Greiner \inst{1}
}

\institute{
      Frankfurt Institute for Advanced Studies, J.-W. Goethe University, 60438 Frankfurt am Main, Germany
\and  Institute for Nuclear Research, Russian Academy of Science, 117312 Moscow, Russia
\and  Kurchatov Institute, Russian Research Center, 123182 Moscow, Russia
}

\abstract{
We study the propagation of nucleons and nuclei in tissue-like media within a Monte 
Carlo Model for Heavy-ion Therapy (MCHIT) based on the GEANT4 toolkit (version 8.2).
The model takes into account fragmentation of projectile nuclei and secondary interactions
of produced nuclear fragments. Model predictions are validated with available experimental 
data obtained for water and PMMA phantoms irradiated by 
monoenergetic carbon-ion beams. The MCHIT model describes well 
(1) the depth-dose distributions in water and PMMA, (2) the doses measured for fragments 
of certain charge, (3) the distributions of positron emitting nuclear fragments produced
by carbon-ion beams, and (4) the energy spectra of secondary neutrons measured at different 
angles to the beam direction. Radial dose profiles for primary nuclei and for different 
projectile fragments are calculated and discussed as possible input for evaluation of 
biological dose distributions. It is shown that at the periphery of the transverse dose 
profile close to the Bragg peak the dose from secondary nuclear fragments is comparable 
to the dose from primary nuclei. 
}

\maketitle

\section{Introduction}
\label{intro}

Radiation therapy of deep-seated tumors with protons and carbon
ions~\cite{Amaldi:Kraft:2005} exploits an inverse energy deposition profile
for such heavy particles with the Bragg peak at the end of their range in tissues. 
In this situation a favourable ratio between the maximum dose in tumor
volume and the minimum dose in healthy tissues 
at the entrance path through the patient's body can be provided. 
An additional advantage of carbon ions stems from their increased 
biological effectiveness  in killing tumor cells 
close to the Bragg peak~\cite{Amaldi:Kraft:2005}. This also helps in sparing healthy tissues 
beyond the tumor.

High dose gradients can be achieved in carbon-ion radiation therapy. This makes possible 
a very selective impact on the tumor, but requires thorough treatment planning.
This should include reliable calculations of doses from secondary nuclear fragments, which are also 
highly ionizing particles with increased biological effect. We argue 
that the GEANT4 toolkit~\cite{GEANT4-Webpage:2006}, 
general-purpose Monte Carlo software for particle transport 
calculations~\cite{Agostinelli:etal:2003,Allison:etal:2006}, can be successfully used    
for quantifying depth and radial dose profiles for primary and secondary nuclei 
of therapeutic energies.

\section{MCHIT model}
\label{MCHIT}

The current version of our Monte Carlo Model for Heavy-ion Therapy 
(MCHIT)~\cite{Pshenichnov:etal:2005,Pshenichnov:etal:2006}  is based on the 
version 8.2 of the GEANT4 toolkit~\cite{GEANT4-Webpage:2006}. Electromagnetic interaction 
of primary and secondary charged particles is described within a set of models
called ``standard electromagnetic physics''~\cite{Agostinelli:etal:2003,Allison:etal:2006}.
Ionization energy loss and straggling of charged particles due to interaction with atomic
electrons are taken into account along with multiple Coulomb scattering on atomic nuclei.
At each simulation step the energy loss of a charged particle is calculated according to
the Bethe-Bloch formula with the mean excitation potential set to 77 eV for water molecules and
68.5 eV for PMMA.   

The binary cascade model is used to describe the collisions of energetic hadrons and nuclei with 
nuclei from the phantom medium~\cite{Agostinelli:etal:2003,Allison:etal:2006}.
Depending on the collision impact parameter, nuclear remnants of various size and excitation 
energy are produced in nucleus-nucleus collisions.  
Evaporation of nucleons, deutrons, tritons, $^{3}$He and $^{4}$He is a dominant de-excitation
process at excitation energies of nuclear remnants below 3 MeV per nucleon. The decay of 
highly excited nuclear remnants is described within the statistical Fermi break-up model. 
This model is used to describe the explosive decay of light hot fragments up to 
$^{16}$O with excitation energies above 3 MeV per nucleon~\cite{Bondorf:etal:1995}. 
The decay of a nuclear system at high excitation 
energies is characterized by a great number of open channels. 
This makes difficult to perform calculations on the basis of analytical 
formulae and approximations for the fragmentation cross sections which are based only on rather 
scarce experimental data. Instead, Monte Carlo models for nuclear 
fragmentation are employed in calculations.  MCHIT predictions are verified with the available 
experimental data on the depth-dose distributions measured in tissue-like phantoms.

In addition to beam particles, electromagnetic and hadronic interactions 
of secondary particles with phantom material are included 
in simulations. More details on the physical models employed in MCHIT can be found 
in refs.~\cite{Pshenichnov:etal:2005,Pshenichnov:etal:2006}

\section{Validation of MCHIT with experimental data}

We consider several characteristics relevant to the transport of a carbon-ion beam in tissue-like
media, which complement each other in validating the MCHIT model.
Since many characteristics are measured with different 
experimental techniques, we consider for the MCHIT validation data collected in 
several experiments.
 
First, the model has confronted with the total depth-dose distributions measured in water and PMMA.
Then, the doses measured for secondary fragments of different charge, e.g. from hydrogen to boron nuclei,
are considered. They give important information on $^{12}$C fragmentation
in violent nucleus-nucleus collisions with large overlap of nuclear densities which create  
hot nuclear systems. The GEANT4 models describing the decay of hot nuclei can be tested with these data. 
On the other hand, the spatial distributions of nuclear fragments close in 
mass to the projectile, e.g. $^{10}$C and $^{11}$C, which are measured via 
their $\beta^{+}$-decay, provide information on peripheral nucleus-nucleus
collisions. Finally, the energy spectra of secondary neutrons measured at various angles bring 
complementary information on the nuclear reactions between projectile and target nuclei.

\subsection{Depth-dose distributions for primary and secondary fragments}

\begin{figure}[t]
\centering
\resizebox{1.08\columnwidth}{!}{%
   \includegraphics{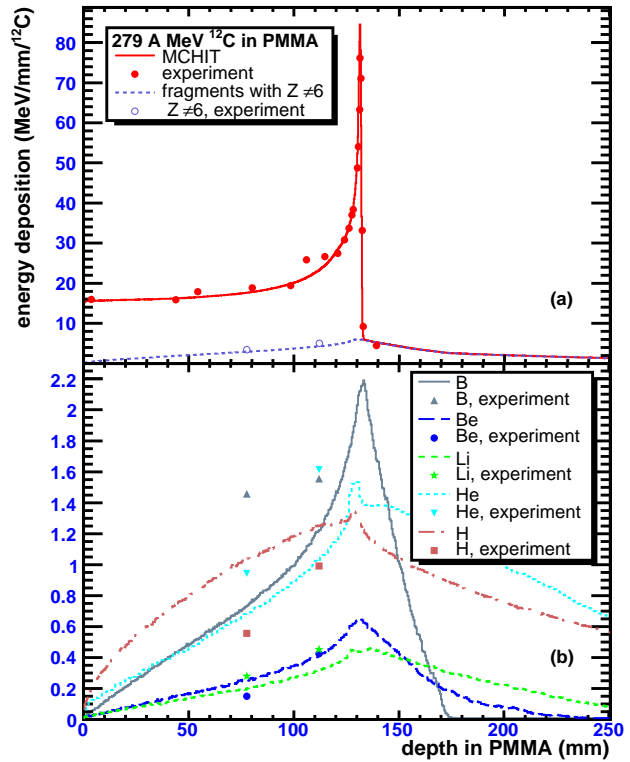}  
}
\caption{Calculated depth-dose distributions for 279 A MeV $^{12}$C beam in PMMA:
(a) the total dose (solid-line histogram) and the dose from fragments other than carbon
(dashed-line histogram);
(b) the doses from hydrogen, helium, lithium, beryllium and boron fragments.  
Experimental data~\cite{Matsufuji:etal:2003} for the doses from fragments of certain
charge are shown by various symbols.}
\label{fig:1}
\end{figure}

The calculated distribution of the total dose in a $15\times 15\times 25$~cm 
polymethylmethacrylate (PMMA) phantom irradiated by 279.23 A MeV $^{12}$C beam
is shown in fig.~\ref{fig:1} and compared with the experimental data by
Matsufuji et al.~\cite{Matsufuji:etal:2003}.
The calculations were performed for a Gaussian beam profile of 10~mm FWHM
by splitting the phantom into thin slices and calculating the 
energy deposited in each of the slices per beam particle. The energy spread of the
beam was assumed to be Gaussian with the FWHM of 0.2\% of the reported beam energy.

The experimental data~\cite{Matsufuji:etal:2003} were normalized to the entrance 
dose and were superimposed on the calculated depth-dose distribution by shifting 
the data points in depth by +3.62~mm. We attribute this correction to possible uncertainties in the 
measurements of the initial beam energy. 

The contributions to the total dose in PMMA phantom associated with different fragments 
were measured at depth of 78 and 112 mm for hydrogen, helium, lithium, beryllium and boron 
fragments~\cite{Matsufuji:etal:2003}.
These fractions were multiplied by the absolute values of calculated dose  
and plotted in fig.~\ref{fig:1}(b) together with MCHIT predictions for each kind of fragment. 
Only secondary fragments, mainly protons and helium nuclei, propagate beyond the distal edge 
of the Bragg peak. As their relative biological effectiveness (RBE)~\cite{Amaldi:Kraft:2005} 
differs from the RBE for primary ions, this should be taken into account in
calculating biological dose beyond the Bragg peak region.

As shown in fig.~\ref{fig:1}, the MCHIT model successfully 
describes the total dose as well as the sum of doses obtained from fragments other than carbon.
The latter accounts for $\sim 20$\% of the total dose at 78 and 112 mm depth.  
The contributions from lithium and  beryllium fragments are also well described, while 
the dose from helium and boron fragments are underpredicted, possibly due to underestimation of their
yields. Further development of the GEANT4 models for nucleus-nucleus collisions 
is required to eliminate this disagreement.

\subsection{Yields of positron-emitting fragments}

Nuclear fragments $^{10}$C and $^{11}$C, which have the same charge as $^{12}$C projectiles,
are abundantly produced in peripheral nucleus-nucleus collisions.
These unstable nuclei are formed via the loss of one or two neutrons and can be detected by 
their $\beta^{+}$-decays accompanied by the positron emission. In addition,
$^{15}$O nuclei are also produced from target $^{16}$O nuclei.
The annihilation of a positron on an electron of the medium results in the 
emission of two gamma quanta, which can be registered by means of the positron emission 
tomography (PET)~\cite{Poenisch:etal:2004,Fiedler:etal:2006}. This method is used
for monitoring of the carbon-ion therapy~\cite{Amaldi:Kraft:2005} by comparison of 
the calculated $\beta^{+}$-activity distribution obtained for the prescribed dose 
with the distribution measured during therapeutic irradiation or soon after it. 
In experiments with water and PMMA phantoms the evolution of
the total $\beta^{+}$-activity profile with time can be  
studied and the contributions from specific $\beta^{+}$-emitting nuclei 
can be extracted~\cite{Fiedler:etal:2006}.

\begin{figure}[t]
\centering
\resizebox{1.05\columnwidth}{!}{%
   \includegraphics{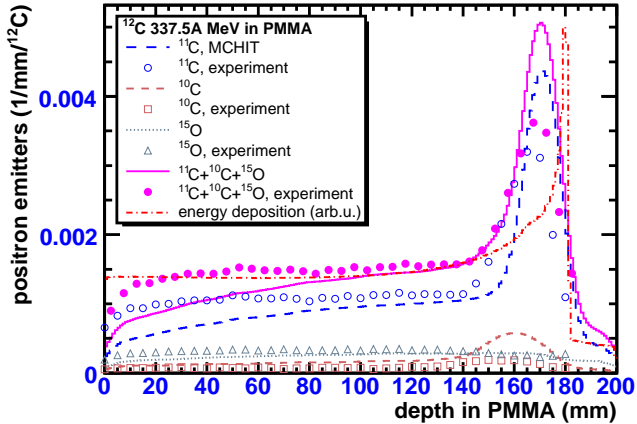}  
}
\caption{Calculated distributions of positron-emitting nuclear fragments, $^{10}$C, 
$^{11}$C and $^{15}$O (long-dashed, dashed and dotted histograms, respectively)
produced by 337.5 A MeV $^{12}$C beam in PMMA. The sum of all contributions is shown by
a solid-line histogram. The corresponding depth-dose distribution is shown by a
dot-dashed histogram. Experimental data obtained at GSI~\cite{Fiedler:etal:2006} 
are shown by points. }
\label{fig:2}
\end{figure}
The depth distributions of positron-emitting nuclei,  $^{10}$C, $^{11}$C  and
$^{15}$O, are shown in fig.~\ref{fig:2} as numbers of these nuclei per unit depth and per
beam particle. The distributions calculated with MCHIT were 
convoluted with a Gaussian of 10~mm FWHM to simulate a finite spatial resolution of a PET scanner.
The depth distribution of $^{11}$C shows clear correlation with the total depth-dose 
distribution, which is also shown in fig.~\ref{fig:2}. 
Although the calculated distributions show more sharp peaks compared to the data~\cite{Fiedler:etal:2006},
the overall agreement is good. The model predicts quite small contributions to the total 
$\beta^{+}$-activity from $^{10}$C and $^{15}$O nuclei, which is in accordance with 
the experimental findings~\cite{Fiedler:etal:2006}.

\subsection{Energy spectra of secondary neutrons}

While most of charged particles deposit their energy locally in the region before and at the Bragg peak,
energetic neutrons produced in collisions of $^{12}$C projectiles with target nuclei 
propagate further at large distances from the tumor volume. Therefore,
it is important to test the MCHIT model with available experimental data for secondary neutrons,
in particular with energy spectra of neutrons measured at different
angles~\cite{Gunzert-Marx:Schardt:Simon:2004}. The estimations of the doses from secondary neutrons 
in proton and heavy ion therapy obtained within the MCHIT model were already given in 
ref.~\cite{Pshenichnov:etal:2005}.

\begin{figure}[t]
\centering
\resizebox{1.08\columnwidth}{!}{%
   \includegraphics{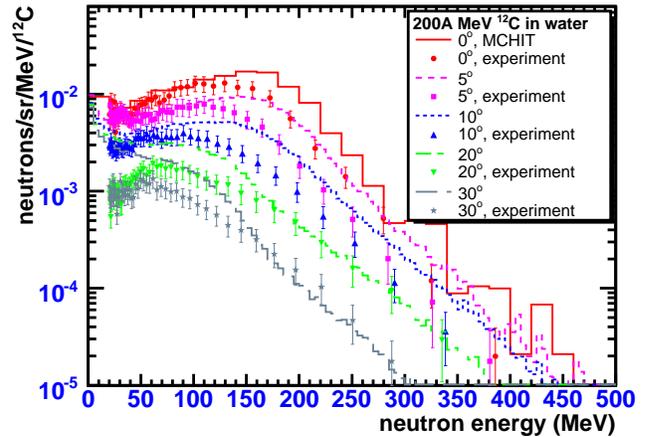}  
}
\caption{Calculated energy spectra of neutrons produced by 200 A MeV $^{12}$C beam in 
a 12.78 cm water phantom (histograms) at  $0^{0}$, $5^{0}$, $10^{0}$, $20^{0}$ and $30^{0}$ 
to the beam axis. Experimental data obtained at 
GSI~\cite{Gunzert-Marx:Schardt:Simon:2004} are shown by points.}
\label{fig:3}
\end{figure}
The calculated energy spectra of secondary neutrons which are
produced by 200 A MeV $^{12}$C beam in a 12.78-cm thick water phantom are shown in fig.~\ref{fig:3}.  
For comparison fig.~\ref{fig:3} also shows the spectra of neutrons measured by 
Gunzert-Marx et al.~\cite{Gunzert-Marx:Schardt:Simon:2004}
at $0^{0}$, $5^{0}$, $10^{0}$, $20^{0}$ and $30^{0}$ to the beam axis behind the water phantom.
As neutrons with energies below 20 MeV were not effectively
registered by detectors, the corresponding data points were excluded from this comparison.
A good overall agreement with data is obtained taking into account the experimental errors in measurements and
three orders of magnitude in variation of neutron yields.
The MCHIT model describes well the yields of neutrons with energies below $\sim$150 MeV 
emitted at $0^{0}$, $5^{0}$ and $10^{0}$ angles, i.e. in forward direction. 
These neutrons were mostly produced in nucleus-nucleus collisions as fragments of primary $^{12}$C nuclei.
Such nuclei were already slowed down due to the ionization energy loss before the collision. 
The kinetic energy of initial beam nuclei is not equally distributed between the emitted neutrons as they
can be emitted at various angles to the beam direction and can have different momenta before the 
collision due to the Fermi motion in the primary nuclei. 
For example, some limited fraction of emitted neutrons has velocities exceeding the beam velocity. 
This is reflected in the exponential fall-off of neutron spectra above the beam energy of 200 MeV. This feature 
is nicely reproduced by the MCHIT model.  

Some discrepancy between the model predictions and the experimental data is found for relatively 
slow neutrons emitted at larger angles $20^{0}$ and $30^{0}$.
Since such neutrons are mostly produced by neutron evaporation, we conclude that  
there is still room for improvement of nuclear de-excitation models used in GEANT4.

\section{MCHIT predictions for dose contributions from nuclear fragments}

\begin{figure}[t]
\centering
\resizebox{1.1\columnwidth}{!}{%
   \includegraphics{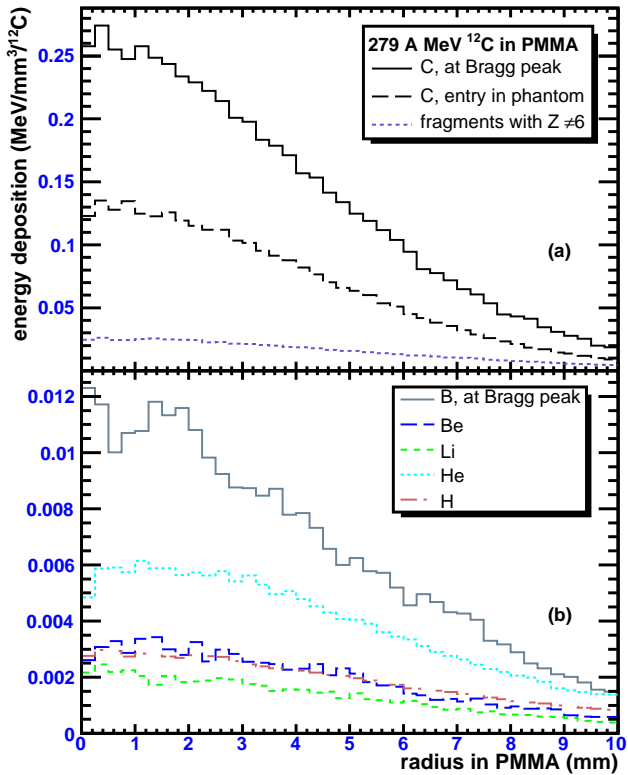}  
}
\caption{Calculated radial dose distributions for 279 A MeV $^{12}$C beam in PMMA:
(a) the dose from carbon ions at the Bragg peak (solid line), at the entry 
(long-dashed line), and the dose from fragments other than carbon
(dashed-line);
(b) the doses from hydrogen, helium, lithium, beryllium and boron fragments.  
}
\label{fig:4}
\end{figure}

Based on the successful validation of the MCHIT model demonstrated above, one can apply it now 
for calculating the spatial dose distributions from therapeutic carbon-ion beam, 
including the contributions from specific secondary 
particles. For example, the radial dose distributions from nuclear fragments with charges from 1 to 6 
calculated for 279 A MeV $^{12}$C beam in PMMA are shown in fig.~\ref{fig:4}. 
The Bragg peak for such energy is located at $\sim 130$~mm. The radial dependence of beam intensity 
in the transverse plane was taken to be Gaussian with FWHM of 10~mm. 
In fig.~\ref{fig:4} the volume energy deposition 
is presented in MeV/mm$^3$ per beam particle, and it is calculated for the depth in 
PMMA from 120 to 130~mm, i.e. in the region 
of the Bragg peak, see fig.~\ref{fig:1}. The radial dose distribution from carbon ions
entering into the phantom (for the depths from 0 to 20~mm) is also
presented in fig.~\ref{fig:4}(a). As shown by calculations, about $\sim 50$\% of primary ions are lost on their
way to the Bragg peak due to nuclear interactions. This reduces the energy deposition from 
carbon ions in the Bragg peak region.
Since only secondary fragments propagate beyond the distal edge of the Bragg peak,
the dose there is delivered exclusively by the projectile fragments, mainly hydrogen and helium nuclei,
as shown in fig.~\ref{fig:1}(b).
 
It is interesting to note that the energy deposition from secondary fragments can not be neglected 
even in the Bragg peak region. There, the contribution to the total dose due to secondary 
fragments is about $\sim 10$\% at the center of the beam profile. 
However, since the secondary fragments produced in nucleus-nucleus collisions can be emitted at large angles, 
their radial distributions are much wider compared to $^{12}$C
beam nuclei, see fig.~\ref{fig:4}(b). Secondary fragments contribute up to $\sim 20$\%
of the total dose at the radii of $\sim 10$~mm.

\section{Discussion}

According to the MCHIT model, the depth-dose distributions are essentially different for 
each kind of secondary fragment, both in longitudinal and transverse directions.
As the biological effectiveness of heavy ions for therapy essentially 
depends on the ion charge, mass and energy, the obtained distributions can be used to calculate 
the depth and lateral profiles of biological doses as well as corresponding cell survival rates. 

The accuracy of the pencil-beam approximation, which assumes that the dose at any depth
and radius is delivered exclusively by primary beam particles, may be valid for proton beams,
where the production of secondary nuclear fragments can be neglected. However, this 
approximation becomes less accurate for carbon-ion beams, where a noticeable fraction of 
dose (up to $\sim 20$\%) is delivered by secondary nuclear fragments. As shown by our  
calculations of the radial dose profiles for each kind of fragments at the Bragg peak, 
the largest contribution apart from $^{12}$C projectiles is obtained from helium and boron fragments.
New experimental studies of radial dose profiles from heavy-ion beams in tissue-like media, 
especially with detection of fragment charge, will be of great importance for the purpose of
validation of the MCHIT model and development of the GEANT4 toolkit.     
In turn, this will contribute significantly to the construction of a realistic theoretical model
for heavy-ion cancer therapy.

\begin{acknowledgement}
This work was partly supported by Siemens Medical Solutions.
We are grateful to Prof. Hermann Requardt and Dr. Thomas Haberer for the discussions 
which stimulated the present study. We are indebted to Prof. Wolfgang Enghardt, 
Fine Fiedler and Dr. Dieter Schardt 
for useful discussions and for providing us the tables of their experimental data.
We thank Dr. Alexander Botvina for discussions concerning fragmentation models.    
\end{acknowledgement}

\end{document}